\documentclass[preprint,12pt]{elsarticle}

\usepackage{amssymb}
\usepackage{amsmath}
\usepackage{nomencl}
\makenomenclature
\journal{Energy Conversion and Management}

\begin{document}

\begin{frontmatter}



\title{How to circumvent the size limitation of
liquid metal batteries due to the Tayler instability}


\author{Frank Stefani\footnote{Corresponding author, E-mail address: F.Stefani@fzd.de}, Tom Weier, Thomas Gundrum, Gunter Gerbeth}

\address{Helmholtz-Zentrum Dresden-Rossendorf, 
P.O. Box 510119 Dresden, D-01314 Dresden, Germany}

\begin{abstract}
Recently, a new type of battery has been proposed that relies
on the principle of self-assembling of a liquid metalloid 
positive electrode,  a liquid electrolyte, and  a
liquid metal negative electrode. While this configuration 
has been claimed to allow arbitrary up-scaling, there
is a size limitation of such a
system due to a current-driven kink-type 
instability that is known as the Tayler instability. 
We characterize this instability in large-scale self-assembled
liquid metal batteries 
and discuss  various technical means how it can be
avoided. 

\end{abstract}


\end{frontmatter}
% \linenumbers

\section{Introduction}
\label{}

With growing deployment of intermittent renewable 
energy sources, such as wind and solar, large 
scale electricity storage becomes an issue of
increasing importance.

In a recent patent application \cite{SADOWAY} 
(see also \cite{BRADWELL} for more details)
a new type of battery has been proposed 
that has the potential to become a key ingredient in 
balancing supply and demand of electrical energy. Its 
working principle relies on
reversible ambipolar electrolysis 
within a self-assembled layered structure of 
a liquid metal, an electrolyte, and
a metalloid. Historically it is worth to note 
that the idea of self-assemblage due to
appropriate density differences was already discussed 
in \cite{AGRUSS}. Such an assembly of three purely 
liquid layers, without the interference of any solid 
phase, allows for the highest 
possible reaction rates known in electrochemistry.
The absence of any vulnerable ceramic electrolyte, as it 
is used in sodium-sulfur (NaS) batteries \cite{NAS}, 
makes the liquid 
metal battery a promising candidate for almost unlimited 
upward scalability, 
which appears to be essential for economic 
competitiveness.

In this paper we will discuss a possible limitation of the
upward scalability of liquid metal batteries 
due to  magnetohydrodynamic (MHD) instabilities
in conducting fluids under the influence of 
externally applied electrical 
currents. The first problem that comes to mind here is 
the interfacial instability that is well known
in aluminium reduction cells
\cite{SELE,DAVIDSON}. This interfacial instability
starts from small  interface deformations between 
cryolite and aluminium which lead to a 
mainly horizontal disturbance current in the 
liquid aluminium. 
The Lorentz force 
resulting from this 
horizontal current in interaction with  a given vertical 
magnetic background field (an unwanted side 
effect of the bus 
bars supplying the current to the cell) drives a horizontal 
motion with a rotating 
interface  that under certain conditions may become unstable.

The onset of this instability depends 
on the height of the cryolite layer which must not fall under 
a certain value.
The usual way to avoid this instability is, therefore, to use
a  thicker electrolyte layer which results in   
a higher electrical resistance. Actually, this resistance
is responsible for the giant Joule losses in aluminium 
reduction cells which are responsible  for  the consumption of  
about 2 per cent of the electricity 
generated world-wide \cite{DAVIDSON}. 
Indeed, this interfacial instability could 
play  a role in liquid batteries, too, and it 
should be considered carefully in order to determine a 
{\it lower} threshold 
for the thickness of the electrolyte layer.

The point of the present paper, however, 
is to highlight and characterize another type of instability
which could set a serious limit for the {\it upper} size of liquid
metal batteries. This instability is well known in astrophysics
under the label Tayler instability (TI) \cite{TAYLER} (sometimes also 
called 
Vandakurov-Tayler instability \cite{VANDAKUROV}).
The TI can be considered as a limiting case of the 
well known 
kink-instability in plasma physics that occurs 
if the so-called
Kruskal-Shafranov stability criterion is violated,
i.e. when the
ratio of axial to azimuthal magnetic field
falls below some critical
value \cite{BERGERSON}.

In our context the TI is a kink-type 
(i.e. non-axisymmetric) instability 
that occurs if the current through a column of
a liquid  metal exceeds some critical value in the
order of kA, depending on the combination 
of the material parameters
conductivity, viscosity and density
\cite{RUEPRE,RUEAN}. 
If this current threshold is 
exceeded, the TI  would lead to a quite vigorous motion and 
very likely to an undesired mixing of the three layers 
which should definitely be avoided for the 
battery to function properly

The main goal of this paper is to  figure out the principle
importance of the TI for liquid metal batteries.
In a very first attempt to quantify the critical current, we 
will use a one-dimensional equation system for the 
hypothetical case of a cylindrical fluid with 
infinite length. This is of course a strong idealization 
of a real battery with its finite ratio of
length to radius
and its possible deviation from the cylindrical shape.
Note, however, that there are indeed visions of 
high capacity batteries for which
the ratio of length to radius of the liquid 
metal and/or metalloids 
might be in the order of one or larger 
\cite{SADOWAY,BRADWELL}.
For any detailed battery design  
the correct determination of the current threshold 
would require expensive two-dimensional simulations 
(in the spirit of \cite{GELLERT}) 
which is beyond the scope of the present paper.

The paper starts with a short outline of the recent ideas 
about liquid metal batteries. It continues with a characterization
of the TI in ideal fluids which will help us to find a most
simple means on how the TI can be avoided. Then, for the 
most widely discussed liquid metals Mg and Na, we will
compute the critical  currents which should be considered 
as conservative estimates for the onset of the TI. 
For those liquids we will also propose a simple means by which 
the TI can be shifted to significantly higher critical 
currents.

\section{Liquid metal batteries}

Liquid metal batteries are thought to play 
a big role in future grid energy storage. 
Sodium-sulfur batteries with
a power around 30\,MW are already used in a Japanese wind-park. 
They work with a cell voltage of around 2\,V, at temperatures of 
around 300$^\circ$C, and they have
AC-AC efficiency of 75-80\% \cite{NAS}.
Their main problem, however, is the use of a ceramic 
$\beta''$- alumina
solid electrolyte which is susceptible to thermal shocks
and which also results in size limitations due to 
the difficulties to manufacture it in 
thin cross sections while maintaining structural strength. 

\begin{figure}[tbh]
  \begin{center}
    \includegraphics[width=8.5cm]{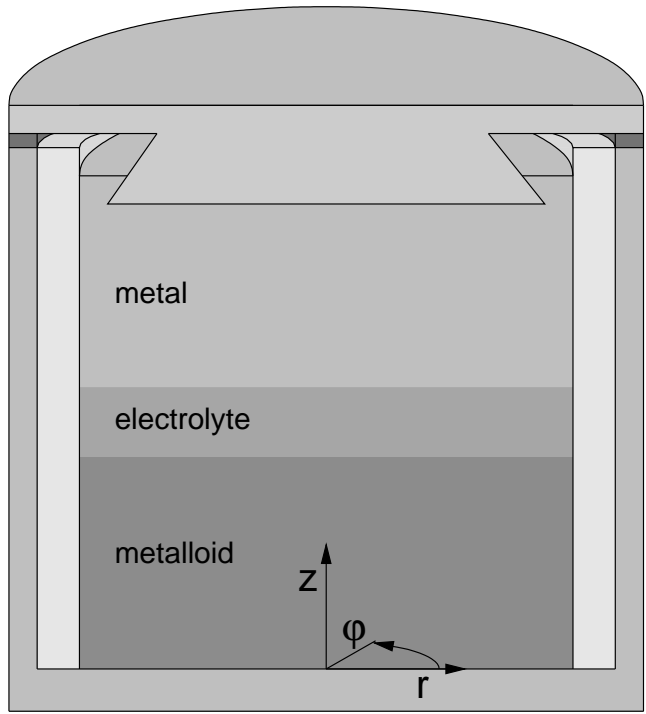}
    \caption{Conceptual design of a liquid metal self-assembling 
    cell with coordinate system.}
    \label{battery1}
  \end{center}
\end{figure}

In order to avoid problems of this sort, 
a perfectly liquid battery was proposed in 
\cite{SADOWAY,BRADWELL}. Such a battery consists 
basically of three liquid layers which are self-assembling 
due to their different densities and their mutual immiscibility
(see Fig. \ref{battery1}). 
The lowest layer, consisting of a metalloid like Sb or Bi, 
represents
the positive electrode of the battery where the negative ions 
of the metalloid are oxidized during the charge process. 
The second layer, the
electrolyte, is a molten salt which is chosen in such a way 
that it is immiscible with both the metal and the metalloid,
allows for the diffusion of their respective ions, is
electronically insulating, and ionize the molten product
of the metal and the metalloid. As potential electrolyte 
components Na$_2$S, Li$_2$S, CaS, Na$_2$Se, Li$_2$Se, and Na$_3$Sb 
are discussed \cite{BRADWELL}. 
% CHANGE
Obviously, the electrolytes' conductivity should be as high as possible
in order to minimize Joule losses. However, unlike in liquid metal
concentration cells \cite{AgrussKaras:1967}, where the electrolytes'
role is merely that of an ion conductor, liquid metal battery 
electrolytes have to absorb a considerable amount of metal and
metalloid ions. This requires high solubility of the metal and
metalloid ions in the electrolyte as well as a relatively large
electrolyte volume. The latter exigence entails thick electrolyte
layers opposing the need for low cell resistance. The use of external
electrolyte reservoirs and a circulation system would be a possible
mean to meet these conflicting goals.
% CHANGE END

The third and top layer 
consists of a molten earth alkaline metal such as Mg 
or, possibly, an alkaline metal such as Na. This layer 
serves as negative electrode where the metal 
cations are reduced during  the charging process. 

It goes without saying that the safe operation of such 
a cell requires the
self-assembled layer structure of the three liquids 
to remain stable during charging and discharging. 

For the quite similar case of aluminium  reduction cells 
the interfacial instability 
is well  known to set some lower limit for the thickness of the
electrolyte below which the fluid starts to undergo a wavy motion
that can ultimately interrupt the electrolysis.
The same sort of instability has to be carefully considered 
for liquid batteries, too.

The focus of this paper is, however, 
on the existence of an upper limit of
the size of the battery beyond which the Tayler instability would
stir up the liquids.

\section{The Tayler instability for the ideal fluid}

The Tayler instability is well known in astrophysics where
it had been first discussed by Vandakurov \cite{VANDAKUROV} and
Tayler \cite{TAYLER}. It can be considered as a limiting case of
the kink-instability
in plasma physics that sets in when the safety factor (basically 
the ratio of axial to azimuthal magnetic field) falls 
below some critical value.

For an ideal fluid, without any viscosity and resistivity, 
Tayler \cite{TAYLER} 
had given the condition
of stability  of a purely azimuthal magnetic field
$B_{\varphi}(r)$ 
\nomenclature{$B_{\varphi}$}{magnetic induction in azimuthal direction
(T)}
with respect to non-axialsymmetric perturbations
(proportional to $\exp{(i \varphi)}$)
\nomenclature{$r$,$\varphi$,$z$}{cylindrical coordinates}
\nomenclature{$i$}{$\sqrt{-1}$}
in the following form:
\begin{eqnarray}
-\frac{d}{d r}(r B^2_{\varphi}(r))>0 \; .
\label{stabil}
\end{eqnarray}

Despite its restriction to ideal fluids, which will be 
relaxed in the next section,
it is worthwhile 
to analyze this expression a bit further. 
With 
view on our goal to avoid the TI in batteries,
we will consider not only the obvious 
case of a 
% CHANGE
%full cylindrical fluid
completely filled cylindrical vessel
% CHANGE END
with a homogeneous current distribution, but
also the more general situation of a homogeneous 
current through the liquid in a 
hollow cylinder, together with  an additional {\it independent} current 
along the axis.

\begin{figure}[tbh]
  \begin{center}
    \includegraphics[width=8.5cm]{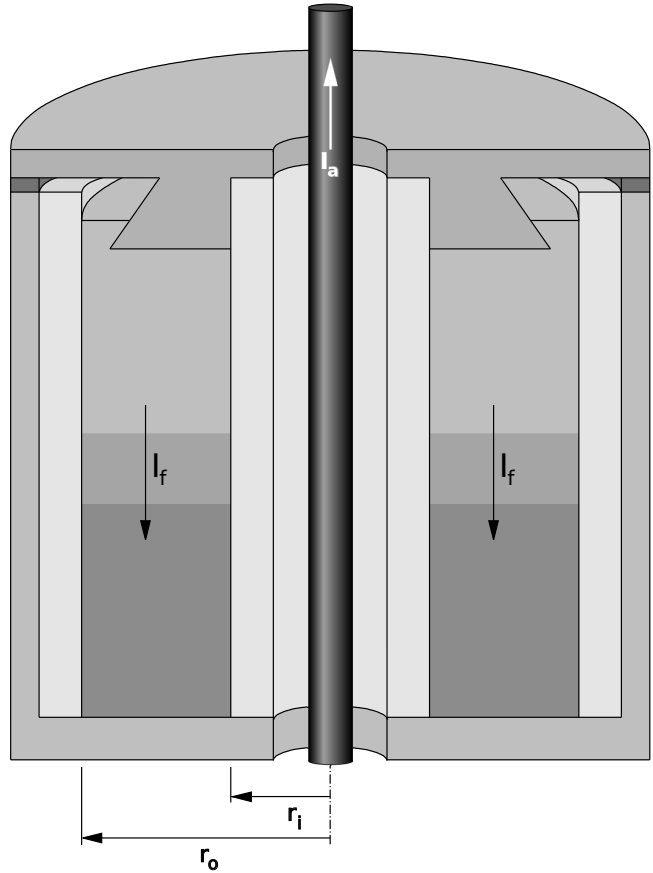}
    \caption{Sketch of a liquid metal battery with a central bore and a
      conductor for a stabilizing axial current.}
    \label{battery2}
  \end{center}
\end{figure}

Assume the fluid to fill the hollow cylinder
between the inner radius $r_{\text{i}}$ and the outer radius 
\nomenclature{$r_{\text{i}}$}{inner radius (m)}
$r_{\text{o}}$, with the ratio of the radii denoted by 
\nomenclature{$r_{\text{o}}$}{outer radius (m)}
$\eta=r_{\text{i}}/r_{\text{o}}$. The current 
\nomenclature{$\eta$}{ratio of inner and outer radius}
density $j$ through this cylindrical gap is assumed 
\nomenclature{$j$}{current density (A/m$^2$)}
homogeneous so that the total 
current through the liquid
is $I_{\text{f}}=\pi (r^2_{\text{o}}-r^2_{\text{i}}) j$. Assume further 
\nomenclature{$I_{\text{f}}$}{total current through the liquid (A)}
the existence of 
an independent 
current $I_{\text{a}}$ along the 
\nomenclature{$I_{\text{a}}$}{independent axial current (A)}
central axis of the cylinder.
Applied to liquid metal batteries, this scheme is depicted in
Fig. \ref{battery2}.
Under these conditions the azimuthal magnetic 
field  in the fluid acquires the form
\begin{eqnarray}
B_{\varphi}(r)=\alpha r+\beta /r \; 
\end{eqnarray}
with the constants $\alpha$ and $\beta$ given by
\begin{eqnarray}
\alpha&=&\frac{\mu_0}{2 \pi} \; \frac{I_{\text{f}}}{r^2_{\text{o}}}  \; \frac{1}{1-\eta^2} \; , \\
\beta&=&\frac{\mu_0}{2 \pi} \; \left[ I_{\text{a}}-I_{\text{f}} \frac{\eta^2}{1-\eta^2}   \right] \; .
\end{eqnarray}
\nomenclature{$\mu_0$}{magnetic permeability of the free space
  (Vs/(Am))}
\nomenclature{$\alpha, \beta$}{constants}
The stability condition (\ref{stabil}) can then be
re-written  in the form
\begin{eqnarray}
\frac{\beta^2}{r^2}-2  \alpha \beta -3 \alpha^2 r^2 >0 \; .
\end{eqnarray}

From the specific radial dependence of this 
expression it is evident that the most 
dangerous position for the instability to occur is
close to $r_{\text{o}}$. Therefore, in order to
identify the critical values for 
$\beta$, 
it is sufficient 
to solve the 
quadratic equation 
\begin{eqnarray}
\beta^2-2 \alpha \beta r^2_{\text{o}}-3 \alpha^2 r^4_{\text{o}} =0  \; .
\end{eqnarray}
only at $r_{\text{o}}$. This quadratic equation has 
two solutions which are $\beta_1=-a r^2_{\text{o}}$ and  $\beta_2=3a r^2_{\text{o}}$. 
Expressed in terms of the
current $I_{\text{a}}$ and $I_{\text{f}}$, this can be
rewritten into the following two
conditions for stability:
\begin{eqnarray}
I_{\text{a}}<-I_{\text{f}}
\end{eqnarray} 
or
\begin{eqnarray}
I_{\text{a}}>I_{\text{f}} \frac{3+\eta^2}{1-\eta^2} \; .
\end{eqnarray}

These two inequalities  are the main result of this section.
They suggest two simple possibilities to avoid the TI.
Either one applies $I_{\text{a}}$ in opposite direction to $I_{\text{f}}$: then 
$I_{\text{a}}$ has to be at 
least as strong as $I_{\text{f}}$. Or 
one applies  $I_{\text{a}}$ in the same 
direction as $I_{\text{f}}$. Then $I_{\text{a}}$ has to be 
factor $(3+\eta^2)/(1-\eta^2)$ stronger than $I_{\text{f}}$. 

\begin{figure}[tbh]
  \begin{center}
    \includegraphics[width=8.5cm]{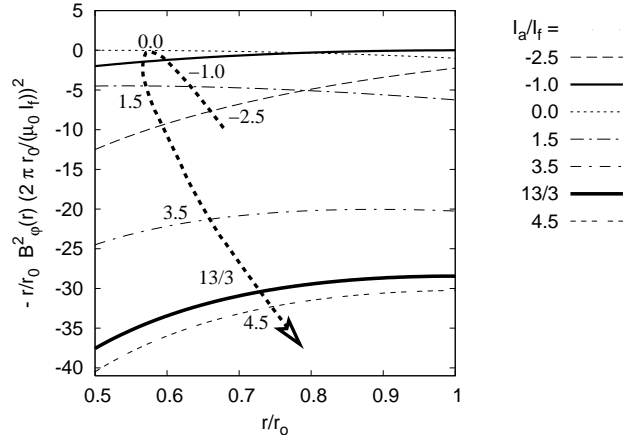}
    \caption{Dependence of the normalized function $-r B^2_{\varphi}(r)$
      on the parameter $I_{\text{a}}/I_{\text{f}}$ for the particular case $\eta=0.5$. 
      Note the monotonic increase of this function both for 
       $I_{\text{a}}/I_{\text{f}}<-1$ and for  $I_{\text{a}}/I_{\text{f}}>13/3$ which indicates the 
       absence of the Tayler instability.
        The thick dashed line with an arrow indicates the 
        change of the profiles
       when going from negative to positive $I_{\text{a}}/I_{\text{f}}$}
    \label{ideal}
  \end{center}
\end{figure}

This is illustrated in Fig. \ref{ideal} where we show, for 
the particular case $\eta=0.5$, 
the (normalized) function $-r B^2_{\varphi}(r)$ for 
various values of $I_{\text{a}}/I_{\text{f}}$. For all curves 
with $-1<I_{\text{a}}/I_{\text{f}}<13/3$ we
can identify regions (at least close to $r_{\text{o}}$) where 
the function $-r B^2_{\varphi}(r)$ has an outward decreasing part.
Both for $I_{\text{a}}/I_{\text{f}}<-1$ and for  
$I_{\text{a}}/I_{\text{f}}>13/3$, however,  the 
function $-r B^2_{\varphi}(r)$ increases monotonically.

For the purpose of batteries, the first version with $I_{\text{a}}<-I_{\text{f}}$ 
seems much more convenient. The limiting case $I_{\text{a}}=-I_{\text{f}}$
can be implemented by simply 
returning the battery current through the inner hole of the battery.

\section{The Tayler instability for real fluids}

While in case of an ideal 
fluid with a homogeneous current distribution
the TI would set in already for arbitrary small 
currents, the main effect of finite  
viscosity and resistivity is to shift 
this threshold to some finite value.

In this section we will determine the critical currents for
some real liquid metals which seem most important in our context, 
and we will study their dependence on the radii ratio $\eta$.

In the appendix we delineate the numerical scheme for the 
determination of the critical current which relies on the scheme 
outlined in \cite{RUEPRE,RUEAN}. In this respect it 
is important to note that the relevant number here is 
the so-called 
Hartmann number $Ha=B R\sqrt{\sigma/(\rho \nu)}$, where
\nomenclature{$Ha$}{Hartmann number}
\nomenclature{$B$}{magnetic induction (T)}
\nomenclature{$\sigma$}{electrical conductivity (S)}
\nomenclature{$\rho$}{density (kg/m$^3$)}
\nomenclature{$\nu$}{kinematic viscosity (m$^2$/s)}
\nomenclature{$R$}{radius of the cell (m)}
$B$ is the magnetic field strength and $R$ is the radius of the 
cell. Hence, the larger
the factor $\sqrt{\sigma/(\rho \nu)}$, the easier the TI will be 
excited for a given current. 
This means. in turn, that if we have a self-assembled layer
of liquids, the most critical one of them is that with the highest
factor. It turns out quickly that the 
typical metals (Mg, Na) forming the negative electrode are more
prone to TI than the half-metals (Sb, Bi) forming the positive electrode.

In the following we will consider two of the most relevant materials. 
The first is Mg at 700$^{\circ}$C with the parameters
$\sigma=3.61 \times 10^6$\,S/m, $\rho=1.54 \times 10^3$\,kg/m$^3$, 
$\nu=6.74 \times 10^{-7}$\,m$^2$/s, the second is Na at 300$^{\circ}$C with
$\sigma=5.99 \times 10^6$\,S/m, $\rho=0.878 \times 10^3$\,kg/m$^3$, 
$\nu=3.94 \times 10^{-7}$\,m$^2$/s.

\begin{figure}[tbh]
  \begin{center}
    \includegraphics[width=8.5cm]{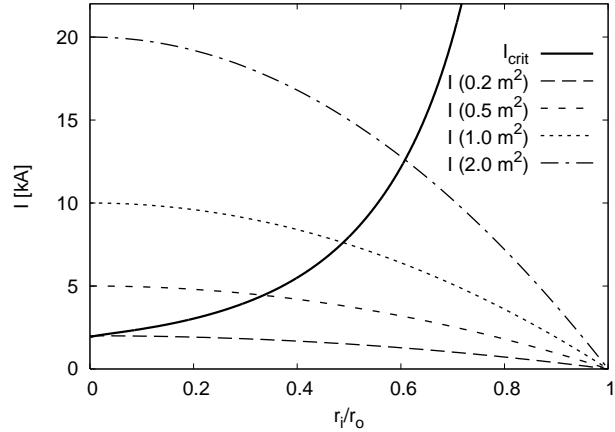}
    \caption{Critical current in dependence on 
    $\eta$ for Mg at 700$^{\circ}$C, 
     under the assumption of a current 
    density of 10 kA/m$^2$.}
    \label{Magnesium}
  \end{center}
\end{figure}

Let us assume now a current density of 10\,kA/m$^2$ as it has been
discussed as a typical achievable value for liquid metal batteries
\cite{SADOWAY}.
For Mg, the solid line in Fig. \ref{Magnesium} 
shows the increasing dependence 
of the critical current
on the ratio $r_{\text{i}}/r_{\text{o}}$. 
For $r_{\text{i}}/r_{\text{o}}=0$ the critical current
is approximately 2\,kA (note that this value is independent of the
radius). This means, in turn, that the maximum area $\pi r^2_{\text{o}}$ 
(for  the 
assumed current density of 10\,kA/m$^2$) is 0.2\,m$^2$. If 
we would like to increase the bearable
current we have to increase $r_{\text{i}}/r_{\text{o}}$. In this case, however, the 
total  current through the cylindrical gap also decreases as 
$\sim (r^2_{\text{o}}-r^2_{\text{i}})$, which is depicted 
by the various non-solid curves in Fig. \ref{Magnesium}.
At the crossing point between these curves with the solid curve one 
can 
identify the maximum current that would be possible for a given
total area $\pi r^2_{\text{o}}$. For example, take the case of a rather large 
area 
$\pi r^2_{\text{o}}=2$\,m$^2$. Evidently, for this case we cannot 
achieve an expected total current 
of 20\,kA which would be a factor 10 higher than the critical current 
at $r_{\text{i}}/r_{\text{o}}=0$. Yet we  can get a total current of 
around 12.5\,kA, if we choose $r_{\text{i}}/r_{\text{o}}=0.61$. 

\begin{figure}[tbh]
  \begin{center}
    \includegraphics[width=8.5cm]{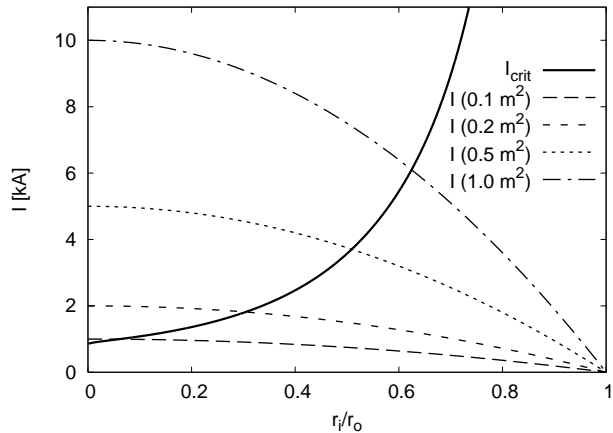}
    \caption{Critical current in dependence 
     on $\eta$ for Na at 300$^{\circ}$C, under the assumption of a current 
    density of 10 kA/m$^2$.}
    \label{Sodium}
  \end{center}
\end{figure}

For sodium the situation is even more critical as shown in
Fig.~\ref{Sodium}. Here, the current threshold at
$r_{\text{i}}/r_{\text{o}}=0$ is approximately 0.86\,kA, and even with an area 
$\pi r^2_{\text{o}}=1$\,m$^2$ we can only achieve a critical current of
6.1\,kA if we go to $r_{\text{i}}/r_{\text{o}}=0.62$

Note that the focus of this section 
was on identifying  the threshold-increasing 
effect of hollow cylinders, without taking into account any
(significant) axial current $I_{\text{a}}$. It is clear, however, that the
idea of the former section to avoid the TI by sending an 
opposite directed current through the middle remains valid also in 
the case of real-world liquid metals.

\section{Conclusion}

The main purpose of this paper was to point out that the TI
could represent a serious obstacle for the upward scalability of liquid
metal batteries. If it occurred in one of the liquid layers
(most likely in the upper metal, having the highest conductivity and
the lowest density) 
it would lead 
to a vigorous motion and very likely to a mixing of the  layers
with possibly ''explosive'' consequences. To give a first conservative
estimate for the critical currents, we have used a simplified 
one-dimensional numerical model
assuming an infinite ratio of height to the radius of a 
battery with
pre-supposed cylindrical geometry. 
For any realistic aspect ratio, this estimate has to be 
concretized by appropriate two-dimensional simulations 
\cite{GELLERT}, 
which would give larger values of the critical currents.
Another, although not a dramatic, modification
of our results should be expected for
the case that the shape of the battery is not a cylinder but, 
e.g., a cuboid.

We have also discussed two ways of how the TI can be avoided. 
One of them is the use of an inner non-conducting tube whose 
mere existence leads to a significant upward shift of the 
critical current. For a typical cell area in the order 
of 1\,m$^2$ this might indeed represent a viable technical means 
for excluding TI. A more 
radical way that can be expected to 
suppress the TI completely is just to return
the current through the hollow cylinder in the middle. 
The same effect could be achieved by applying a parallel 
axial current which is at least a factor 
$\frac{3+\eta^2}{1-\eta^2}$ as large as the current in the fluid.

Many theoretical aspects of the TI in liquid metal batteries 
are yet to be discussed. First experiments are presently 
carried out to study the TI in a cylindrical liquid metal and
to validate various 
possible ways how it can be avoided.

\section*{Acknowledgments}
This work was supported by the Deutsche
Forschungsgemeinschaft under
grant number STE 991/1-1 and 
in the framework of 
SFB 609. Stimulating discussions
with Marcus Gellert, Rainer Hollerbach, J\={a}nis Priede, 
and G\"unther R\"udiger on the Tayler instability are gratefully
acknowledged. We would like to thank Frank-Peter Wei\ss\
for his encouragement and support.

\printnomenclature

%A third method
%relies on the application of an axial magnetic field which could 
%conveniently be produced by the current itself.
%The problem with the latter is that the needed vertical field 
%might would help the interfacial instability to set in.

\newpage
\appendix

\section{The mathematics of Tayler instability}

In this appendix we describe how to 
determine the critical current of the TI
for the hypothetical case of a cylinder with infinite length. 
For this purpose we use a simplified version of the procedure 
described in detail in \cite{RUEPRE}.
We start with the Navier-Stokes equation 
for the velocity field $\bf{V}$ 
\begin{eqnarray}
\frac{\partial {\bf V}}{\partial t}+({\bf V} \cdot {\bf \nabla}){ \bf V}&=&
-\frac{1}{\rho} {\bf \nabla} P+\nu \Delta {\bf V}+\frac{1}{\mu_0} 
({\bf \nabla} \times {\bf B}) \times  {\bf B}
\end{eqnarray}
and the induction equation for the magnetic field $\bf B$
\begin{eqnarray}
\frac{\partial {\bf B}}{\partial t}&=&
{\bf \nabla} \times ({\bf V} \times {\bf B})
+\frac{1}{\mu_0 \sigma} \Delta {\bf B} \; .
\end{eqnarray}
Here, $\rho$ is the density of the fluid, $\nu$ its kinematic viscosity,
$\sigma$ its electrical conductivity and $\mu_0$ the 
magnetic permeability of the free space.
We also note that both fields have to be divergence-free
(we assume incompressibility of the fluid):
\begin{eqnarray}
{\bf \nabla} \cdot {\bf V}=0, \; \;\; {\bf \nabla} \cdot {\bf B}=0 \; .
\end{eqnarray}
Both ${\bf V}$ and ${\bf B}$, as well as the pressure $P$, 
are now split into the basic state
and a perturbation according to
\begin{eqnarray}
{\bf V}= {\bf V}_0+{\tilde {\bf v}} \; \;\; P= P_0+{\tilde p} \; \;\;
{\bf B}= {\bf B}_0+{\tilde {\bf b}} \; .
\end{eqnarray}
In our particular case the basic state of the velocity is just 
${\bf V}_0=0$, while
the basic state of the magnetic field is given by
\begin{eqnarray}
B_{r,0}=0, \;\;\; B_{\varphi ,0}(r)=\alpha r+\beta /r, \;\;\; B_{z,0}=0
\end{eqnarray}
with the constants $\alpha$ and $\beta$ given by
\begin{eqnarray}
\alpha&=&\frac{\mu_0}{2 \pi} \; \frac{I_{\text{f}}}{r^2_{\text{o}}}  \; \frac{1}{1-\eta^2}\\
\beta&=&\frac{\mu_0}{2 \pi} \; \left[ I_{\text{a}}-I_{\text{f}} \frac{\eta^2}{1-\eta^2}   \right] \; .
\end{eqnarray}
We employ now the usual normal mode analysis, by searching for 
solutions of the linearized equations for 
\begin{eqnarray}
{\tilde {\bf v}}(r,\varphi,z,t)={\bf v}(r)\exp(i(k z+m \varphi+\omega t))\\
{\tilde p}(r,\varphi,z,t)=p(r)\exp(i(k z+m \varphi+\omega t))\\
{\tilde {\bf b}}(r,\varphi,z,t)={\bf b}(r)\exp(i(k z+m \varphi+\omega t))
\end{eqnarray}
This way, we end up with an eigenvalue system of ordinary 
differential equations (in $r$): 
\begin{eqnarray}
\frac{d v_r}{d r}+\frac{v_r}{r}+\frac{im}{r} v_{\varphi}+ik v_z&=&0\\
\frac{d p}{d r}+i\frac{m}{r} X_2+ik X_3+\left(k^2+\frac{m^2}{r^2} \right) u_r&&\nonumber \\
-i {\rm Ha_{\text{i}}}^2\frac{m}{r} B_{\varphi,0} b_r+2 {\rm Ha_{\text{i}}}^2 \frac{B_{\varphi}}{r} b_{\varphi}&=&0\\
\frac{d X_2}{d r}-\left(k^2+\frac{m^2}{r^2} \right) v_{\varphi}+2i\frac{m}{r^2} v_r&&\nonumber \\
+\frac{{\rm Ha_{\text{i}}}^2}{r}\frac{d}{dr}(rB_{\varphi ,0}) b_r+i {\rm Ha_{\text{i}}}^2 \frac{m}{r} B_{\varphi ,0} b_{\varphi}
-i\frac{m}{r} p&=&0\\
\frac{d X_3}{d r}+\frac{X_3}{r}-\left(k^2+\frac{m^2}{r^2} \right) v_z
-ikp
+i{\rm Ha_{\text{i}}}^2 \frac{m}{r} B_{\varphi ,0} b_z&=&0\\
\frac{d b_r}{d r}+\frac{b_r}{r}+\frac{im}{r} b_{\varphi}+ik b_z&=&0\\
\frac{d b_z}{d r}-\frac{i}{k} \left(k^2+\frac{m^2}{r^2} \right) b_r
+\frac{m}{kr} X_4-\frac{m}{kr} B_{\varphi ,0} u_r&=&0\\
\frac{d X_4}{d r}-\left(k^2+\frac{m^2}{r^2} \right)b_{\varphi}+i\frac{2m}{r^2} b_r&&\nonumber\\
-r\frac{d}{dr} \left(\frac{B_{\varphi ,0}}{r} \right) v_r +i\frac{m}{r} B_{\varphi ,0} v_{\varphi}&=&0
\end{eqnarray}
where the abbreviations
\begin{eqnarray}
X_2=\frac{d v_{\varphi}}{dr}+\frac{v_{\varphi}}{r},\;\;\;
X_3=\frac{d v_z}{dr},\;\;\; X_4=\frac{d b_{\varphi}}{dr}+\frac{b_{\varphi}}{r}\;\;\;
\end{eqnarray}
have been used. The governing parameter ${\rm Ha_{\text{i}}}$ here is the so-called Hartmann 
number, defined with the
magnetic field at the inner radius according to 
\begin{eqnarray}
{\rm Ha_{\text{i}}}&=&B_{\varphi}(r_{\text{i}}) \sqrt{r_{\text{i}}(r_{\text{o}}-r_{\text{i}})} \sqrt{\sigma/(\rho \nu)}\\
&=&\frac{\mu_0 I_{\text{a}}}{2 \pi r_{\text{i}}} \sqrt{r_{\text{i}}(r_{\text{o}}-r_{\text{i}})} \sqrt{\sigma/(\rho \nu)} \; .
\end{eqnarray}
In the numerical implementation we will always use a (possibly
very small) axial current.
The boundary conditions, both at $r_{\text{i}}$ and $r_{\text{o}}$, for the 
velocity are the 
usual no-slip conditions
\begin{eqnarray}
v_r=v_{\varphi}=v_z=0
\end{eqnarray}
As for the magnetic field, we assume insulating boundary conditions 
which can be written as
\begin{eqnarray}
b_{\varphi}-\frac{m}{kr_{\text{i}}} b_z&=&0\\
b_r+\frac{i}{I_m(kr_{\text{i}})} \left( \frac{m}{kr_{\text{i}}} 
I_m(kr_{\text{i}})-I_{m+1}(kr_{\text{i}}))  \right) b_z&=&0
\end{eqnarray}
at the inner boundary $r_{\text{i}}$ and 
\begin{eqnarray}
b_{\varphi}-\frac{m}{kr_{\text{o}}} b_z&=&0\\
b_r+\frac{i}{K_m(kr_{\text{o}})}\left(\frac{m}{kr_{\text{o}}}K_m(kr_{\text{o}})-K_{m+1}(kr_{\text{o}}))  
\right) b_z&=&0
\end{eqnarray}
at the outer boundary $r_{\text{o}}$. Note that $I_m$ and $K_m$ represent 
the modified Bessel functions.

In order to determine the critical currents for a given ratio
of $\eta$ and $I_{\text{a}}/I_{\text{f}}$ we solve the above eigenvalue system 
by means of a shooting method adopted from \cite{RECIPES}. Then we have
to look for those wavenumbers $k$ that lead to a minimum 
critical $I_{\text{a}}$.


%% References


%\bibliographystyle{elsarticle-num}

%% Authors are advised to submit their bibtex database files. They are80 (2010


\end{document}